\documentclass[fleqn,usenatbib,useAMS]{mnras}

\usepackage{newtxtext,newtxmath}

\usepackage[T1]{fontenc}
\usepackage{ae,aecompl}

\usepackage{graphicx}	
\usepackage{amsmath}	
\usepackage{amssymb}	
\usepackage{color}
\usepackage[]{threeparttable}
\usepackage{subcaption}
\definecolor{coralie}{rgb}{1.0, 0.50, 0.75}
\definecolor{james}{rgb}{0.48, 0.40, 0.93}

\definecolor{vero}{rgb}{0.875, 0.257, 0.859}

\usepackage{soul}
\definecolor{jaune}{rgb}{1.0, 1.0, 0.0}

\title[MOBSTER - I. First-light observations of known magnetic stars]{Magnetic OB[A] Stars with \textit{TESS}: probing their Evolutionary and Rotational properties (MOBSTER) - I. First-light observations of known magnetic B and A stars}

\author[A. David-Uraz et al.]{A. David-Uraz,$^{1}$\thanks{E-mail: adu@udel.edu} C. Neiner,$^{2}$ J. Sikora,$^{3,4}$ D.~M. Bowman,$^{5}$ V. Petit,$^{1}$\newauthor S. Chowdhury,$^{6}$ G. Handler,$^{6}$ M. Pergeorelis,$^{1}$ M. Cantiello,$^{7,8}$ D.~H. Cohen,$^{9}$\newauthor C. Erba,$^{1}$ Z. Keszthelyi,$^{4,3}$ V. Khalack,$^{10}$ O. Kobzar,$^{10}$ O. Kochukhov,$^{11}$\newauthor J. Labadie-Bartz,$^{12}$ C.~C. Lovekin,$^{13}$ R. MacInnis,$^{1}$ S.~P. Owocki,$^{1}$ H. Pablo,$^{14}$\newauthor M.~E. Shultz,$^{1}$ A. ud-Doula,$^{15}$
G.~A. Wade,$^{4}$ and the MOBSTER Collaboration
\\
$^{1}$Department of Physics and Astronomy, University of Delaware, Newark, DE 19716, USA\\
$^{2}$LESIA, Observatoire de Paris, PSL University, CNRS, Sorbonne Universit\'{e}, Univ. Paris Diderot, Sorbonne Paris Cit\'{e}, 5 place\\ Jules Janssen, F-92195 Meudon, France\\
$^{3}$Department of Physics, Engineering Physics \& Astronomy, Queen's University, Kingston, ON K7L 3N6, Canada\\
$^{4}$Department of Physics and Space Physics, Royal Military College of Canada, PO Box 17000 Kingston, ON K7K 7B4, Canada\\
$^{5}$Instituut voor Sterrenkunde, KU Leuven, Celestijnenlaan 200D, B-3001 Leuven, Belgium \\
$^{6}$Nicolaus Copernicus Astronomical Center, Bartycka 18, 00-716 Warszawa, Poland \\
$^{7}$Center for Computational Astrophysics, Flatiron Institute, 162 5th Avenue, New York, NY 10010, USA \\
$^{8}$Department of Astrophysical Sciences, Princeton University, Princeton, NJ 08544, USA \\
$^{9}$Department of Physics and Astronomy, Swarthmore College, Swarthmore, PA 19081, USA \\
$^{10}$D\'{e}partement de Physique et d'Astronomie, Universit\'{e} de Moncton, Moncton, NB E1A 3E9, Canada \\
$^{11}$Department of Physics and Astronomy, Uppsala University, Box 516, 75120, Uppsala, Sweden \\
$^{12}$Instituto de Astronomia, Geof\'{i}sica e Ci\^{e}ncias Atmosf\'{e}ricas, Universidade de S\~{a}o Paulo, Rua do Mat\~{a}o 1226, Cidade\\ Universit\'{a}ria, 05508-900 S\~{a}o Paulo, SP, Brazil \\
$^{13}$Physics Department, Mount Allison University, Sackville, NB, E4L 1E6, Canada\\
$^{14}$AAVSO Headquarters, 49 Bay State Rd., Cambridge, MA, 02138, USA \\
$^{15}$Penn State Scranton, 120 Ridge View Drive, Dunmore, PA 18512, USA
}

\date{Accepted 2019 April 22. Received 2019 April 22; in original form 2019 March 14}

\pubyear{2019}

\begin{document}
\label{firstpage}
\pagerange{\pageref{firstpage}--\pageref{lastpage}}
\maketitle

\begin{abstract}
In this paper we introduce the MOBSTER collaboration and lay out its scientific goals. We present first results based on the analysis of nineteen previously known magnetic O, B and A stars observed in 2-minute cadence in sectors 1 and 2 of the Transiting Exoplanet Survey Satellite (\textit{TESS}) mission. We derive precise rotational periods from the newly obtained light curves and compare them to previously published values. We also discuss the overall photometric phenomenology of the known magnetic massive and intermediate-mass stars and propose an observational strategy to augment this population by taking advantage of the high-quality observations produced by \textit{TESS}.
\end{abstract}




\section{Introduction}

Unlike their lower-mass counterparts (e.g. \citealt{2007ApJ...657..486B}), massive and intermediate-mass stars are not known to host magnetic fields generated by contemporaneous dynamos at their surfaces. Instead, there exists a distinct population of magnetic O-, B- and A-type stars whose fields appear to be of fossil origin \citep{1982ARA&A..20..191B, 2015IAUS..305...61N, 2017arXiv170510650A}. Magnetic OBA stars host strong surface fields that typically have large-scale, mostly dipolar topologies that are stable over decades (for a broader review, see e.g. \citealt{2009ARA&A..47..333D}). Despite the fact that this spectral type range covers a large variety of stellar parameters (masses, radii, luminosities), the incidence rate of detectable magnetic fields in these stars seems to be uniformly small, $\sim$10\%, and the field properties show no systematic change with, e.g., mass or luminosity \citep{2014IAUS..302..265W, 2015IAUS..307..342M, 2017MNRAS.465.2432G, 2019MNRAS.483.2300S}. 

These stars exhibit a range of phenomenologies that can be understood in terms of the interaction between their magnetic fields and their photospheres or atmospheres. In particular, since the magnetic field is not generally aligned with the rotational axis, many observable quantities across the electromagnetic spectrum are found to be rotationally modulated, in a manner that is understood in the context of the Oblique Rotator Model \citep{1950MNRAS.110..395S}.

In earlier-type (OB) magnetic massive stars, the interaction between the magnetic field and the strong, supersonic line-driven stellar winds leads to the formation of an observable \textit{magnetosphere} \citep{2002ApJ...576..413U}, the global characteristics of which are determined by the stellar rotation rate, wind parameters and magnetic field strength \citep{2005MNRAS.357..251T, 2009MNRAS.392.1022U, 2013MNRAS.429..398P}. More specifically, \textit{dynamical} magnetospheres (DMs) are formed around slow rotators and consist of outflowing ionized material channeled by the closed magnetic field lines creating strong shocks near the magnetic equator. The shocked gas cools radiatively and falls back onto the stellar surface in complex dynamic flows. Fast rotators additionally form a \textit{centrifugal} magnetosphere (CM): centrifugal support of plasma above the co-rotation radius prevents material from falling back onto the star, leading to the formation of dense clouds which co-rotate with the stellar surface.

The presence of a magnetosphere around these massive stars is inferred from observations at many different wavelengths. In the optical, periodic variations in H$\alpha$ are detected (e.g. \citealt{2007MNRAS.381..433H, 2011AJ....141..169B, 2012MNRAS.426.2208G, 2013MNRAS.429..177R}), as well as line profile variations in wind-sensitive resonance lines (primarily in the ultraviolet; e.g. \citealt{1996A&A...312..539S, 2012MNRAS.422.2314M}), variable emission in the X-rays \citep{2004A&A...421..715S, 2005ApJ...628..986G} and in the infrared \citep{2015A&A...578A.112O}, and, in some cases, emission in the radio is also observed (e.g. \citealt{2015MNRAS.452.1245C, 2017MNRAS.465.2160K, 2018MNRAS.476..562L}). Some of these observations can be understood using magnetohydrodynamic (MHD) simulations 
(e.g. \citealt{2013MNRAS.428.2723U, 2013MNRAS.431.2253M, 2014ApJS..215...10N}), as well as through simplified analytic prescriptions such as the ``Analytic Dynamical Magnetosphere" model for DMs (ADM; \citealt{2016MNRAS.462.3830O}) and the ``Rigidly Rotating Magnetosphere" model for CMs (RRM; \citealt{2005MNRAS.357..251T}).

In intermediate-mass stars (late B-type and A-type stars), the effect of magnetism is somewhat different, as they do not possess the same fast, dense winds as higher mass stars. Instead, fossil fields are known to affect diffusive processes in their radiative envelope \citep{1970ApJ...160..641M, 2007A&A...475..659A}, often leading to chemical abundance patches on their surface and, as a result, rotational modulation of spectral lines of various chemical elements in their spectra \citep[e.g.][]{Kochukhov2015,Yakunin2015,Silvester2017}. The energy distributions associated with magnetic intermediate-mass stars are also known to exhibit abnormal flux depressions in the UV \citep{Kodaira1969,Adelman1975,Maitzen1976}. This is understood to be a direct consequence of the presence of strong surface magnetic fields \citep{Kochukhov2005a}.

The Transiting Exoplanet Survey Satellite (\textit{TESS}; \citealt{2015JATIS...1a4003R}) is the latest high-precision photometric space mission. During its two-year nominal mission time, it will observe 85 per cent of the full sky, in overlapping sectors of 96$\times$24 deg, for a total of roughly 470 million point sources observed in full-frame images. The targets will be observed with various temporal baselines, from 27.4\,d to almost 1\,y, depending on their position on the sky. Designed to search for exoplanets transiting in front of their host star, \textit{TESS} was launched on April 18, 2018, and started delivering public data of the first two observed 27.4-d sectors of the sky in December 2018. Given their high quality, these data can be used for a wide range of astronomical investigations in addition to exoplanet detections, including detailed asteroseismology and, in our case, studies of rotational modulation associated with stellar magnetism in early-type stars.

Of particular relevance to \textit{TESS}, both massive and intermediate-mass stars with surface magnetic fields are known to exhibit periodic photometric variations associated with rotational modulation. In the case of earlier-type stars (earlier than about B1), these variations are due to a changing column density as the viewing angle through the magnetosphere varies with phase. This phenomenon is observed for instance in the light curve of HD 191612 \citep{2011MNRAS.416.3160W}. 
For later-type magnetic stars (later than about B5), photometric variations are associated with chemical inhomogeneities on their surface manifesting as brightness spots due to flux redistribution (e.g. \citealt{1970ApJ...161..685P}). In fact, two classes of variable stars, as defined by the General Catalog of Variable Stars (GCVS; \citealt{2017ARep...61...80S}), correspond to chemically peculiar B and A-type stars understood to host strong magnetic fields, the SX Arietis and $\alpha^2$ Canum Venaticorum ($\alpha^{2}$\,CVn) variables, respectively. Finally, for stars with an intermediate spectral type (roughly between B1 and B5), their photometric variability can be caused by either one of the two aforementioned effects, although it is often dominated by photospheric inhomogeneities. Indeed, even when magnetospheric eclipses are seen (as is the case with $\sigma$~Ori~E; \citealt{2008MNRAS.389..559T, 2013ApJ...769...33T}), there can still be a signature associated with chemical spots (as shown for that same star; \citealt{2015MNRAS.451.2015O}). Furthermore, pulsations also contribute to photometric variability in a subset of magnetic OBA stars, notably slowly-pulsating B stars (SPB; \citealt{1985A&A...152....6W}), $\beta$ Cephei variables (e.g. \citealt{1952AnAp...15..157S}), $\delta$ Scuti pulsators (e.g. \citealt{1937LicOB..18...77F}) and rapidly oscillating Ap (roAp) stars \citep{1982MNRAS.200..807K}. 

Therefore, optical time-series photometry can be a powerful means of characterizing known magnetic OBA stars and identifying promising magnetic candidates. Moreover, in some cases additional phenomena that are associated with photometric signatures, such as binarity and pulsations, can be modelled to provide valuable constraints on the physical parameters of these stars. With the rise of high-precision space-based photometric missions over the last decade (e.g. \textit{MOST}, CoRoT, \textit{Kepler}, K2, BRITE-Constellation) and given 
the comparatively limited availability and time-sampling of high-resolution spectropolarimetry, we can leverage 
the observations acquired by these missions to further our understanding of phenomena associated with stellar magnetism in the upper Hertzsprung-Russell Diagram (HRD). 
This approach has been employed to infer the characteristics of putative magnetic fields in extra-Galactic Of?p stars (\citealt{2015A&A...577A.107N}; \citealt{2018CoSka..48..149M}), a spectral class known to be associated with magnetism in the Milky Way \citep{2017MNRAS.465.2432G}. 

\subsection{The MOBSTER collaboration}

The MOBSTER collaboration (Magnetic OB[A] Stars with \textit{TESS}: probing their Evolutionary and Rotational properties) aims to leverage \textit{TESS} observations to gain further insight and understanding into the nature of magnetic massive and intermediate-mass stars. In particular, this project focuses on three types of targets:

\begin{itemize}

\item Known magnetic OBA stars with rotational periods shorter than $\sim$ 27\,d: these targets will be observed for at least one full rotational cycle during the \textit{TESS} mission and their photometric variations can be used to test and calibrate models by comparing them to synthetic light curves;

\item Known magnetic OBA stars with rotational periods greater than $\sim$ 27\,d: these targets will likely be observed for only part of their rotational cycle; therefore shorter-term, potentially stochastic processes (e.g. dynamic flows in a DM) as well as other phenomena (such as pulsations) can be investigated; and

\item Magnetic OBA candidates: we can identify magnetic candidates directly from their light curves and flag them for spectropolarimetric follow-up; such an observational strategy has proven highly successful for K2 data \citep{2018MNRAS.478.2777B}. Such efforts are led in parallel with similar studies classifying variability types in OBA stars observed by \textit{TESS} (e.g. \citealt{2019ApJ...872L...9P, 2019MNRAS.485.3457B}; Sikora et al., submitted;
for a complementary study of Ap stars using \textit{TESS} data see Cunha et al., submitted).


\end{itemize}

High-quality light curves from \textit{TESS} will allow us to pursue various science goals, such as the asteroseismic characterization of magnetic massive stars (which can teach us about the internal effects of surface magnetic fields; e.g. \citealt{2012MNRAS.427..483B, 2018A&A...616A.148B}) and precise determinations of their rotational periods. In this first paper of a series, we focus on the latter objective as we present newly-determined rotational periods for nineteen known magnetic B and A\footnote{We include in this sample a well-known star exhibiting the classic Ap phenomenology, HD 213637, but which is classified as F1. Given the nature of its peculiarities and the typical uncertainty on the spectral type of chemically peculiar stars, we chose to include it in our sample and consider it for all intents and purposes as an A-type star.} stars which were observed in \textit{TESS} sectors 1 and 2\footnote{While our sample comprises magnetic targets that were observed in sectors 1 and 2, a few of these stars (namely HD 53921, HD 24188 and HD 54118) were also observed in sector 3, so we included these data in our analysis.} and compare them to previously published values. By `known magnetic stars', we mean stars whose surface magnetic fields have been directly diagnosed using the Zeeman effect, in either spectroscopic or spectropolarimetric observations (and sometimes both). The relevant references for each star's magnetic detection can be found in section~\ref{sec:notes}. Seven more B- and A-type stars that were observed in sectors 1 and 2 (HD 10840, HD 19400, HD 58448, HD 65950, HD 221507, HD 203932 and HD 221760) yielded spurious or marginal detections in spectropolarimetric observations, and were not included in this sample.

In Section~\ref{sec:obs}, we present the observations. In Section~\ref{sec:res}, we report results for known magnetic B and A stars, and discuss each star individually. Finally, we present our conclusions and consider future work in Section~\ref{sec:concl}.

\section{\textit{TESS} observations}\label{sec:obs}

The \textit{TESS} data included and analyzed here are the 2-min light curves provided by the \textit{TESS} Science Team and which are publicly available via the Mikulski Archive for Space Telescopes (MAST)\footnote{\url{https://archive.stsci.edu/missions-and-data/transiting-exoplanet-survey-satellite-tess}}. A description of the data processing pipeline of these light curves is provided by \citet{2016SPIE.9913E..3EJ}. 
Full-frame images with 30-min cadence are not considered within the scope of this paper. 
Sector 1 was observed from July 25 to August 22, 2018, while sector 2 was observed from August 23 to September 20, 2018.

To select our sample, we cross-matched the list of observed stars in sectors 1 and 2 with the SIMBAD database \citep{2000A&AS..143....9W} and considered all the stars with known spectral types of A and earlier. 
The orbital period of \textit{TESS} is 13.7 days and a gap is present in the data with this period (i.e. in the middle of each 27.4-d light-curve). 
The instrumental magnitudes of each target are found in the \textit{TESS} Input Catalog (TIC; \citealt{2018AJ....156..102S}). These values are included in Table~\ref{tab:targets}.

Among the $\sim$1150 OBA stars (excluding subdwarfs) available in \textit{TESS} sectors 1 and 2, there are no known magnetic O stars, four known magnetic B stars, and fifteen known magnetic A stars (including HD 213637). Our detrended 2-min light curves of these known magnetic stars are presented in Fig.~\ref{fig:bstars} (B stars) and Fig.~\ref{fig:astars} (A stars). These nineteen objects are discussed below.

\section{Results}\label{sec:res}

\subsection{Photometric analysis}

First, to measure accurate rotational periods using the \textit{TESS} light curves, we improve the detrending of these light curves by manually removing obvious outliers and by fitting rotational modulation (when appropriate) with a periodic model which takes into account both a fundamental rotational frequency and its significant harmonics
, following the approach detailed by \citet{2018A&A...616A..77B}. This allows us to both evaluate the rotational periods precisely and model and subtract the residuals (after having evaluated their scatter) using a locally weighted scatterplot smoothing (LoWeSS; \citealt{doi:10.1080/01621459.1979.10481038, seabold2010statsmodels}) filter, thus improving the long-term detrending.

Rotational periods are reported in Table~\ref{tab:targets}. The uncertainties are propagated using the following equation (from \citealt{1976fats.book.....B} and \citealt{1999DSSN...13...28M}):

\begin{equation}\label{eq:eq}
\sigma_f = \frac{\sqrt{6}\sigma_r}{\pi \sqrt{N}A\Delta t}
\end{equation}

\noindent where $\sigma_f$ and $\sigma_r$ represent, respectively, the uncertainty on the measured frequency and the photometric uncertainty, $N$ is the number of points in the light curve, $A$ is the amplitude of the corresponding peak in the periodogram and $\Delta t$ corresponds to the temporal baseline of the time series. 
To evaluate the photometric uncertainty, we use the scatter of the residuals of the pre-whitened light curve after the multi-harmonic rotational model has been removed, as described above. In our results, we quote uncertainties in measured rotational periods as three times the formal uncertainty determined using Eq.~\ref{eq:eq}, to account for the average difference in noise levels in the high and low frequency regimes (i.e. `pink noise') often found in oversampled time-series data (see \citealt{2003ASPC..292..383S, 2009A&A...506..111D, 2018MNRAS.476..601H, 2018A&A...616A..77B}).

\begin{figure*}
	\includegraphics[width=6.8in]{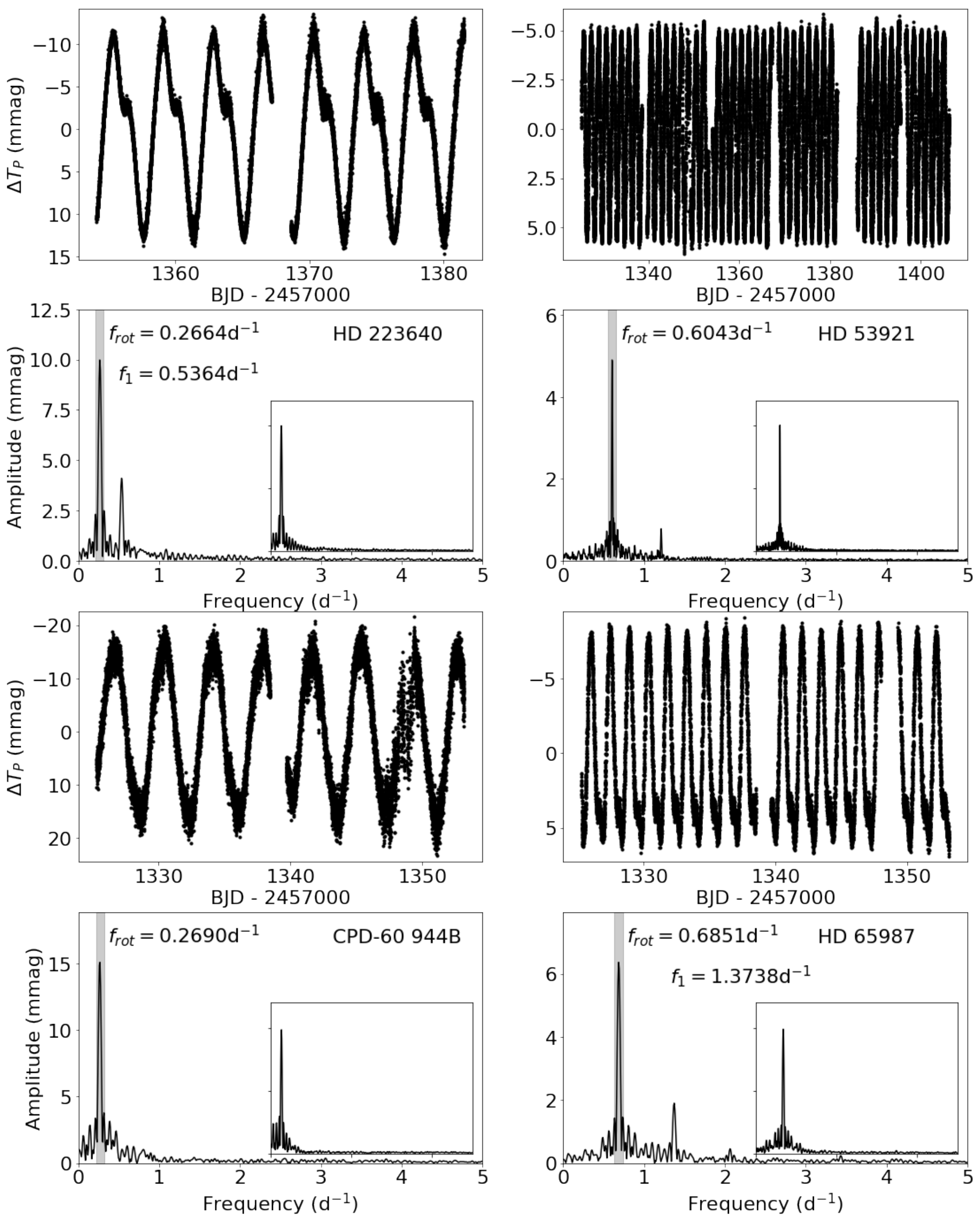}
    \caption{Light curves (top of each panel) and Lomb Scargle periodograms (bottom of each panel) for each of the four known magnetic B stars observed by \textit{TESS} in its sectors 1 and 2. The inferred rotational frequency for each star is highlighted in light grey and labelled, and the first harmonic is also labelled (when applicable). 
    An inset is included to show the window function around the frequency peak with the largest amplitude.}
    \label{fig:bstars}
\end{figure*}

Finally, we perform a Lomb-Scargle analysis \citep{1976Ap&SS..39..447L, 1982ApJ...263..835S} of these detrended light curves of our nineteen targets by using the \texttt{LombScargle} class in the \texttt{astropy.stats} \textsc{Python} package. The results are presented in Figs.~\ref{fig:bstars} and \ref{fig:astars}. These figures present the detrended light curves, together with their Lomb-Scargle periodogram in which the rotational frequency (and its first harmonic) have been identified, when applicable. 

Out of our nineteen targets, thirteen show a clear peak in their periodogram that we ascribe to rotation. For those targets, we show in Figs.~\ref{fig:bstars} and \ref{fig:astars} a `window function' which was obtained by computing a Lomb-Scargle periodogram for a pure sinusoidal signal with a frequency corresponding to the frequency peak with the largest amplitude in the observational periodogram and with the same time sampling as the \textit{TESS} light curve. Some of the light curves that are not considered to show convincing rotational modulation have at least one low-frequency peak (e.g. HD 66318), but those peaks are weak (with amplitudes well below 0.5 mmag) and do not convincingly phase the observations.

Out of the thirteen targets exhibiting signs of rotational modulation, twelve show at least one harmonic of that frequency, the lone exception being CPD-60 944B. The frequencies of the first harmonic peaks are labelled in Figs.~\ref{fig:bstars} and \ref{fig:astars} to confirm their identification. It should be noted that these frequencies are directly measured from the periodograms and are hence much less precise than the period determinations presented in Table~\ref{tab:targets}; hence, they do indeed correspond, within errors, to twice the fundamental frequency.

The amplitudes of the various signals that we detect are well within the typical range associated to $\alpha^2$ CVn variables, with HD 65712 showing the greatest amplitude (nearly 60 mmag). We further detail our findings for individual stars in the following subsection. The light curves of the stars for which a low-frequency peak potentially associated with rotation is recovered are phase folded using our measured rotational periods and presented in the Appendix (Fig.~\ref{fig:app}).



\begin{table*}
\resizebox{\linewidth}{!}{
\begin{threeparttable}
\centering
\caption{
List of known magnetic B- and A-type stars observed by \textit{TESS} in sectors 1 and 2. For each star, we provide the TIC number, a more common identifier (mostly the HD number), the spectral type, the \textit{TESS} instrumental magnitude
, as well as three periods that are measured for the star (and that are long enough to potentially be associated with rotation, although in some cases they may be related to other phenomena such as pulsations): published periods based, respectively, on photometry and on other measurements, and our new refined periods derived from the \textit{TESS} data. 
For these periods, the number appearing between parentheses corresponds to the uncertainty on the final digit of the reported period; the absence of such a number denotes an unreported uncertainty. Finally, we report the largest measured longitudinal magnetic field for each star (and its uncertainty) and the total number of spectropolarimetric observations referenced in the literature (an asterisk denotes a star for which these measurements are considered to be phase resolved, meaning that the sampling in rotational phase is sufficient to trace the variation over the full rotational cycle, while a question mark indicates a lower limit, provided that some sources did not clearly state the total number of observations).}
\label{tab:targets}
\begin{tabular}{@{}rllccccr@{}lc@{}} 
\hline
TIC no. & Name & Sp. type & \textit{T} 
& Pub. period (phot.) & Pub. period (other) & \textit{TESS} period & \multicolumn{2}{c}{$|B_{ \textrm{z, max}}|$} & $N_\textrm{s}$\\
&  &  & (mag) 
& (d) & (d) & (d) & \multicolumn{2}{c}{(G)} \\
\hline
89545031 & HD 223640 & B9pSiSrCr$^a$ & 5.32 
& 3.72251(97)$^c$ & 3.735239(24)$^f$ & 3.7349(5) & 1390$\pm$&220$^{f}$ & $40^*$\\
279511712 & HD 53921 & B9III+B8V$^b$ & 5.80 
& 1.6520$^d$ & 1.6518$^g$ & 1.65183(2) & 511$\pm$&83$^{n}$ & 5 \\
358467049 & CPD-60 944B & B9pSi$^a$ & 8.74 
& 3.7367(3)$^e$ & - & 3.759(2) & 463$\pm$&72$^{n}$ & 2 \\
372913684 & HD 65987 & B9pSiSr$^a$ & 7.65 
& 1.45610(15)$^c$ & $\sim$ 9$^h$ & 1.4561(1) & 738$\pm$&122$^{n}$ & 2 \\
\hline
32035258 & HD 24188 & A0pSi$^a$ & 6.42 
& 2.23047(4)$^c$ & - & 2.23024(4) & 538$\pm$&44$^{n}$ & 1 \\
69855370 & HD 213637 & F1pEuSr$^a$ & 9.18 
& - & - & - & 856$\pm$&45$^{n}$ & 30 \\
139191168 & HD 217522 & A5pSrEuCr$^a$ & 7.17 
& - & - & - & 1123$\pm$&82$^{o}$ & 97 \\
159834975 & HD 203006 & A2pCrEuSr$^a$ & 4.90 
& 2.12230(9)$^c$ & 2.1219$^i$ & 2.122(4) & 650$\pm$&147$^{p}$ & 14 \\
235007556 & HD 221006 & A0pSi$^a$ & 5.82 
& 2.31206(36)$^c$ & 2.31483(40)$^j$ & 2.3119(1) & 990$\pm$&180$^{l}$ & 3 \\
237336864 & HD 218495 & A2pEuSr$^a$ & 9.23 
& 4.2006(1)$^c$ & - & 4.183(6) & 1169$\pm$&56$^{n}$ & 5 \\
262956098 & HD 3988 & A0pCrEuSr$^a$ & 8.10 
& - & - & - &  -&  & 0\\
277688819 & HD 208217 & A0pSrEuCr$^a$ & 7.09 
& 8.3200(84)$^c$ & 8.44475(11)$^k$ & 8.317(1) & 1843$\pm$&214$^{q}$ & 20?* \\
278804454 & HD 212385 & A3pSrEuCr$^a$ & 6.78 
& 2.5062(2)$^c$ & - & 2.5062(2) & 639$\pm$&40$^{n}$ & 2 \\
279573219 & HD 54118 & A0pSi$^a$ & 5.30 
& 3.2724(10)$^c$ & 3.27533(20)$^l$ & 3.2759(2) & 1680$\pm$&120$^{l}$ & 15* \\
280051011 & HD 18610 & A2pCrEuSr$^a$ & 8.06 
& - & - & - &  -&  & 0 \\
281668790 & HD 3980 & A7pSrEuCr$^a$ & 5.70 
& 3.9517(1)$^c$ & 3.9516(3)$^m$ & 3.951(3) & 1688$\pm$&29$^{n}$ & 5 \\
348717688 & HD 19918 & A5pSrEuCr$^a$ & 9.12 
& - & - & - & 777$\pm$&109$^{r}$ & 2? \\
358467700 & HD 65712 & A0pSiCr$^a$ & 9.30 
& 1.94639(54)$^c$ & - & 1.9460(2) & 1296$\pm$&71$^{n}$ & 2 \\
410451752 & HD 66318 & A0pEuCrSr$^a$ & 9.56 
& 0.77688(52)$^c$ & - & - & 6480$\pm$&91$^{n}$ & 2 \\
\hline
\end{tabular}
\begin{tablenotes}
\small
\item References for the spectral types, published periods and maximum longitudinal fields are the following: \textit{a}) \citet{2009A&A...498..961R}; \textit{b}) \citet{1984ApJS...55..657C}; \textit{c}) Cunha et al. (submitted); \textit{d}) \citet{2011MNRAS.414.2602D}; \textit{e}) \citet{2015A&A...581A.138B};  \textit{f}) \citet{1992A&A...258..389N}; \textit{g}) \citet{1999A&A...343..872A}; \textit{h}) \citet{1972ApJ...172..355A}; \textit{i}) \citet{1974A&A....32...21M}; \textit{j}) \citet{1995A&A...294..223L}; \textit{k}) \citet{1997A&A...320..497M}; \textit{l}) \citet{1993A&A...269..355B}; \textit{m}) \citet{1980A&A....81..323M}; \textit{n}) \citet{2015A&A...583A.115B}; \textit{o}) \citet{2004A&A...415..661H}; \textit{p}) \citet{1958ApJS....3..141B}; \textit{q}) \citet{2017A&A...601A..14M}; and \textit{r}) \citet{2006A&A...450..763K}.
\end{tablenotes}
\end{threeparttable}
}
\end{table*}

\begin{figure*}
	\includegraphics[width=6.8in]{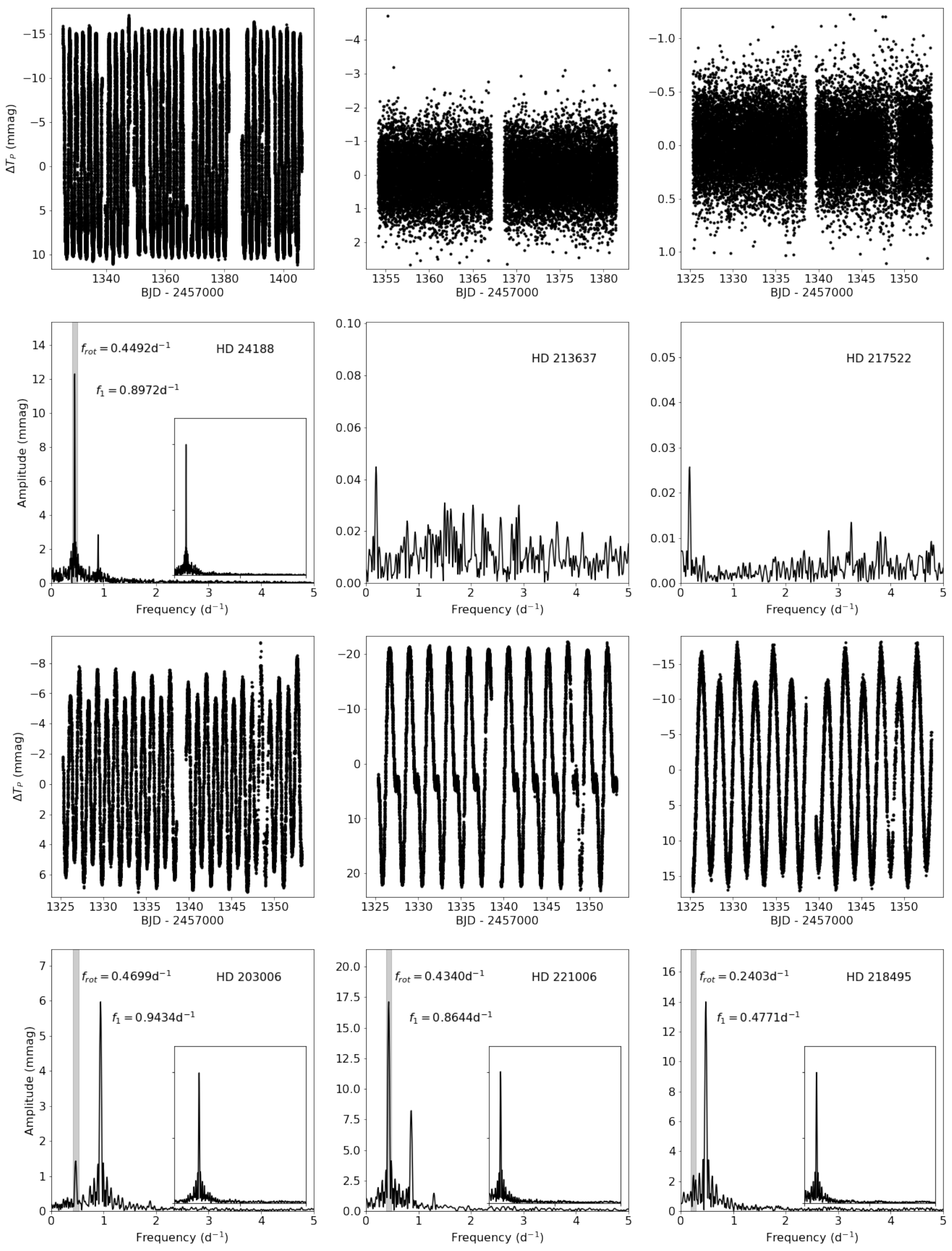}
    \caption{Same as Fig.~\ref{fig:bstars}, but for the A stars. For some periodograms, there is no convincing signature of rotational modulation; in those cases we did not include any window function.
    }
    \label{fig:astars}
\end{figure*}
\begin{figure*}
\ContinuedFloat
	\includegraphics[width=6.8in]{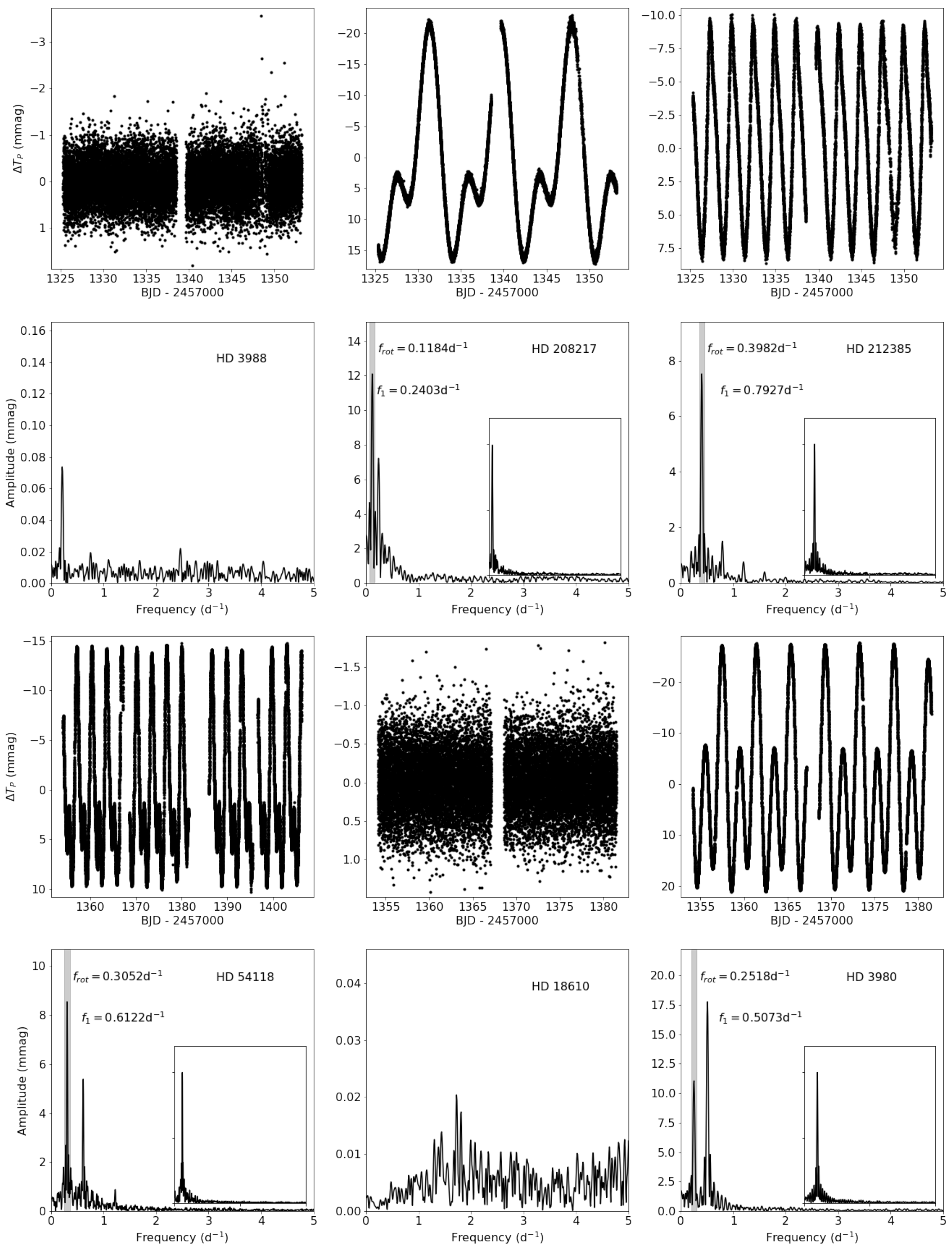}
    \caption{\textit{Continued}}
\end{figure*}
\begin{figure*}
\ContinuedFloat
	\includegraphics[width=6.8in]{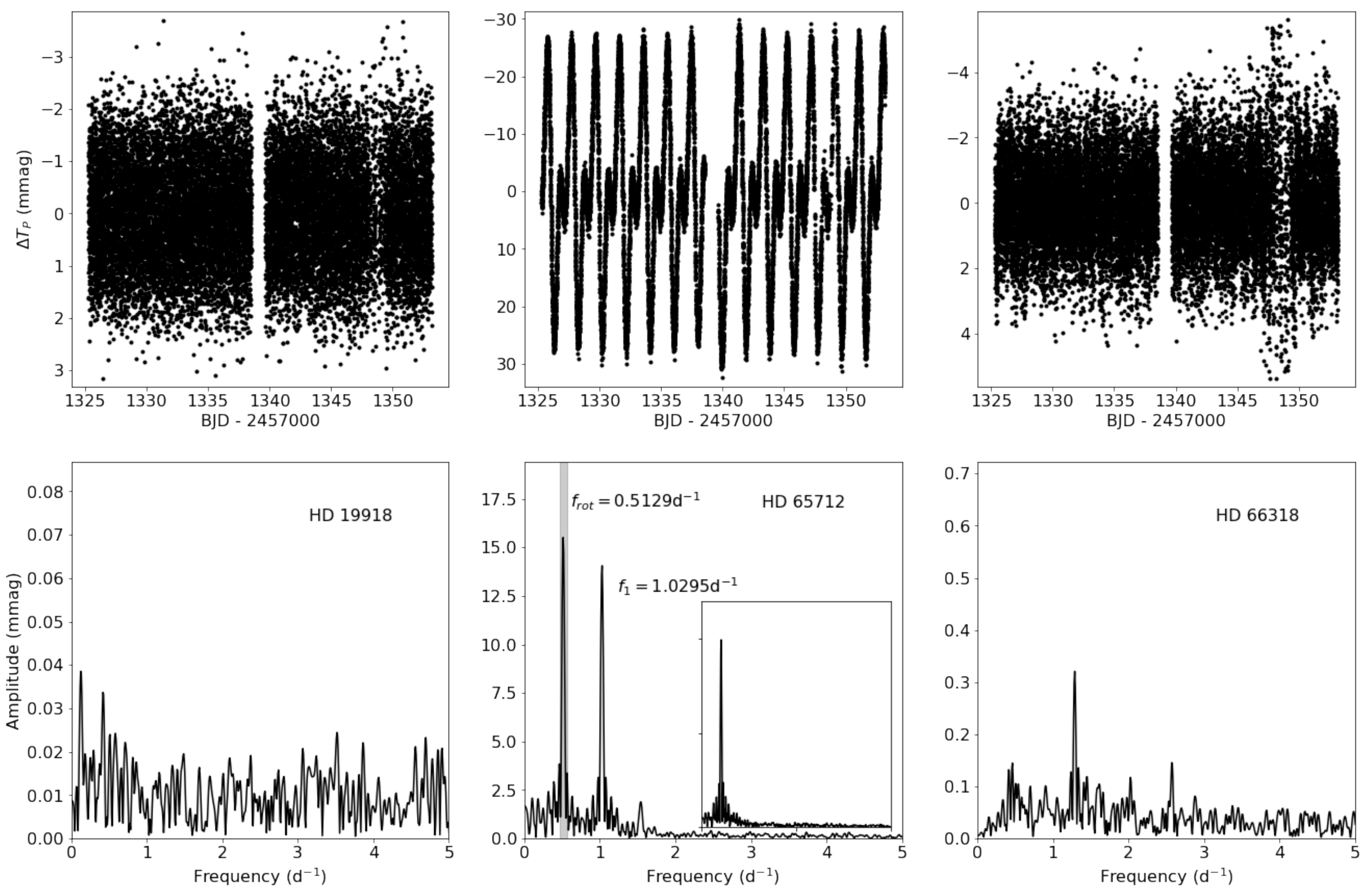}
    \caption{\textit{Continued}}
\end{figure*}

\subsection{Notes on individual stars}\label{sec:notes}

Most of the following stars have multiple period determinations in the literature, therefore we report the most recent one prior to our own study in Table~\ref{tab:targets}. For a number of our targets, an independent period analysis was carried out using the \textit{TESS} data by Cunha et al. (submitted); we report their periods for comparison purposes whenever possible (only in Table~\ref{tab:targets} for concision, except in the case of HD 218495 since theirs is the only prior rotational period determination available in the literature).\\

\noindent\textit{TIC 89545031 (= HD 223640, B9pSiSrCr, V = 5.18)}\\

A photometric period of $3.73 \pm 0.03$\,d was derived for HD 223640 \citep{1971PASP...83..474M}, a period that was also confirmed spectroscopically \citep{1974A&A....34...53M, 1975A&A....39..263M}, with infrared photometry \citep{1991A&A...248..179C, 1998A&AS..131...63C}, and with variations of He\textsc{i} $\lambda$5876 (that were however observed to be out of phase with the other reported variability; \citealt{1999A&AS..134..211C}).
The photometric period has been refined a number of times since \citep{1973A&A....26..385K, 1991IBVS.3635....1N}, and more recently to $3.735239 \pm 0.000024$\,d using photometry and magnetic measurements \citep{1992A&A...258..389N}. Changes in the shape of the light curve over time have been proposed to be indicative of precession of the magnetic axis \citep{1994A&AS..103....1A, 1997A&AS..125...65A, 1999BaltA...8..369A}. 
This star was also observed with Hipparcos, and an automated period search yielded a value of 3.7342\,d \citep{2011MNRAS.414.2602D}. We measure a period of 3.7349$\pm$0.0005\,d in the \textit{TESS} data, consistent with the previous determination; however, we cannot evaluate the precession hypothesis given the different bandpasses involved. 

HD 223640 was first found to be magnetic by \citet{1958ApJS....3..141B}, although its field was not detected in a more recent FORS1 spectropolarimetric observation \citep{2006AN....327..289H, 2015A&A...583A.115B}. This is not surprising as the observation was taken at the rotational phase corresponding to a magnetic null in the field curve of \citet{1992A&A...258..389N}.\\

\noindent\textit{TIC 279511712 (= HD 53921, B9III + B8V, V = 5.64)}\\

Out of the nineteen known magnetic stars discussed in this study, HD 53921 is the only one that is not known to exhibit chemical peculiarities in its spectrum. 
The Hipparcos photometry reveals a 1.65-d period which was attributed to a gravity mode pulsation, implying that HD 53921 is an SPB star \citep{1998A&A...330..215W}. This period determination was refined to 1.6520\,d using the same dataset, but applying an automatic period search \citep{2011MNRAS.414.2602D}. Its radial velocity is also found to vary with the same period (1.6518\,d; \citealt{1999A&A...343..872A}). In addition, lower frequencies were detected in the radial velocity measurements, possibly indicating long-period orbital variations (\citealt{2000A&A...355.1015D}; consistent with a visual double with separation > 1 arcsec, e.g. \citealt{2001AJ....121.1597H}). This scenario was confirmed by \citet{2002A&A...393..965D}, who found HD 53921 to be an eccentric SB1 with a period of $\sim$340\,d.
The components have a $\sim$1 magnitude difference in the $V$ band, with the brightest having a spectral type of B9III, and the fainter one B8V \citep{1984ApJS...55..657C}.  It should be noted that both components lie in the same \textit{TESS} pixel.

The photometric variations are difficult to reconcile with non-radial pulsations \citep{2002MNRAS.330..855T} and mode identification has proven to be arduous \citep{2005A&A...432.1013D}. The detected frequency lies within the (large) range of possible rotational frequencies based on its projected rotational velocity and inferred radius \citep{2015MNRAS.450.1585S}, and could therefore rather be the rotational period.

Although a first spectropolarimetric observation of HD 53921 yielded only a marginal detection \citep{2006AN....327..289H}, it was confirmed to be magnetic \linebreak by \citet{2012A&A...538A.129B}. \citet{2015A&A...583A.115B} later identified the B9III primary as the magnetic star.
The published frequency is recovered in the \textit{TESS} data (1.65183$\pm$0.00002\,d) with a harmonic pattern typical of rotational modulation. We thus propose that this frequency does not correspond to pulsations, but rather to rotation. Further magnetic characterization would help confirm the nature of this frequency.\\

\noindent\textit{TIC 358467049 (= CPD-60 944B, B9pSi, V = 8.76)}\\

CPD-60 944B is a member of a $\sim$10" visual pair within the open cluster NGC 2516 \citep{1975PASP...87..721S}. Its visual companion, CPD-60 944A, might itself be a binary \citep{2000AJ....119.2296G}, a hypothesis that has also been supported by \citet{2014MNRAS.443.1523G}, who propose an orbital period of 121.6\,d or 182.5\,d (they also propose CPD-60 944B to be a HgMn star, a claim that is not supported by other studies). There has historically been some degree of confusion between both of these stars, and the membership of CPD-60 944B in NGC 2516 has been questioned (e.g. \citealt{2008AJ....136..118F}).

\citet{2015A&A...581A.138B} measured a photometric period of 3.7367$\pm$0.0003\,d and attributed it to component A, although both visual components were within the aperture of ASAS-3.  
We report this period in Table~\ref{tab:targets}, since we find a similar period in the \textit{TESS} observations (3.759$\pm$0.002\,d). However both visual components also fall within the \textit{TESS} aperture. 
A marginal magnetic field detection was reported by \citet{2006A&A...450..777B}, and later confirmed by \citet{2015A&A...583A.115B}.\\

\noindent\textit{TIC 372913684 (= HD 65987, B9pSiSr, V = 7.59)}\\

HD 65987 was proposed to be an eclipsing binary system \citep{1975PASP...87..721S}, and has often been classified as an eclipsing Algol variable (e.g. \citealt{2013AN....334..860A}). 
Based on radial velocity measurements, \citet{1972ApJ...172..355A} find a possible periodicity of $\sim$9\,d that could be associated to binarity.
\citet{2014MNRAS.444.1982A} evaluate this system to be a detached main sequence binary, based on its well-established membership in the NGC 2615 cluster (e.g. \citealt{2007A&A...470..685L}). 

Low-level photometric variability was first detected with a putative 1.41-d period \citep{1982IBVS.2103....1N}, which was later refined to 1.44962 $\pm$ 0.00018\,d by \citet{1984A&AS...55..259N}. If this value is assumed to be related to rotation, the implied small value of $R \sin i$ \citep{1991A&A...249..401H} would indicate that the inclination of the rotational axis is likely small. 
The TESS period (1.4561$\pm$0.0001\,d) is similar to that of \citet{1984A&AS...55..259N}, and the data are of much better quality than the phased photometry presented by \citet{1987A&AS...70...33H}. 
While we ascribe this period to rotation, a binary origin of the light curve variations is not conclusively ruled out. Interestingly, a weak peak appears in the periodogram of the non-detrended light curve at a period of about 9 days, corroborating the idea that that period might be linked to binarity.

HD 65987 was found to be magnetic by \citet{2006A&A...450..777B}, and that detection was later confirmed by \citet{2015A&A...583A.115B}.
However, new magnetic measurements with improved phase coverage do not seem to vary according to the photometric period (Landstreet et al., in prep.), making this system harder to interpret. More work will be required to confirm the origin of HD 65987's photometric variability.\\
\\









\noindent\textit{TIC 32035258 (= HD 24188, A0pSi, V = 6.26)}\\

HD 24188 was found to exhibit non-linear proper motion, making it an astrometric binary candidate \citep{2005AJ....129.2420M, 2007A&A...464..377F} even though it is not observed to be a spectroscopic binary \citep{1999A&AS..135..503G}.
Hipparcos photometry revealed a clear periodicity \citep[2.230 d;][]{1998A&AS..133....1P, 2011MNRAS.414.2602D}, which was associated with rotation. We confirm and refine this period with the \textit{TESS} light curve (2.23024$\pm$0.00004\,d). 

Based on a single spectropolarimetric observation, HD~24188 was found to be magnetic by \citet{2006A&A...450..763K}, a conclusion which was since supported through the reanalysis of the same observation \citep{2006AN....327..289H, 2015A&A...583A.115B}.\\ 

\noindent\textit{TIC 69855370 (= HD 213637, F1pEuSr, V = 9.58)}\\

HD 213637 is a roAp star with an 11-min pulsation period \citep{1997IBVS.4507....1M}. It has two known frequencies, which vary in amplitude over time \citep{1998A&A...334..606M}, possibly due to rotation, although no rotational period has been reported in the literature. It is one of the coolest known roAp stars and is likely evolved, and it exhibits magnetically split lines \citep{2003A&A...404..669K}. We do not find any convincing signature of rotational variation in the \textit{TESS} light curve.

This star's magnetic field was further diagnosed using FORS1 observations \citep{2004A&A...415..685H, 2015A&A...583A.115B}. There is no indication of multiplicity \citep{2012A&A...545A..38S}, and based on magnetic field modulus measurements, the rotational period is likely longer than 115\,d \citep{2017A&A...601A..14M}. \\







\noindent\textit{TIC 139191168 (= HD 217522, A5pSrEuCr, V = 7.52)}\\

HD 217522 is a well-studied roAp star (13.72-min pulsational period; \citealt{1983MNRAS.205....3K}). It is one of the few roAp stars for which mode switching has been observed \citep{1991MNRAS.250..477K}.
However, no sign of rotational modulation of the pulsational amplitudes has been found by \citet{2012MNRAS.426..969V}, who conclude that either the stellar surface is not very spotted or the rotational axis is nearly aligned with our line of sight, not allowing us to see rotational modulation, a hypothesis further supported by the very low measured value of $v \sin i$ (3 km s$^{-1}$; \citealt{2015MNRAS.446.1347M}). Additionally, the magnetic axis could be aligned with the rotational axis. While the amplitude modulation occurs on short timescales ($\sim$1\,d), it appears to be stochastic, similar to solar-type oscillations.
This also appears to be qualitatively consistent with the fact that we do not find a significant low-frequency series of harmonics in the \textit{TESS} observations of HD 217522, and in particular, no sign of rotation. Analysis of the high frequency variability using \textit{TESS} observations was performed by Cunha et al. (submitted).

A marginal detection of a magnetic field was reported by \citet{1997A&AS..124..475M}, and has since then been confirmed \citep{2004A&A...415..685H, 2015A&A...583A.115B}. The field strength is not found to vary on the pulsational timescale (\citealt{2004A&A...415..661H}; they acquired 91 short observations within $< 0.25$\,d, a time span insufficient to detect rotational modulation).\\





\noindent\textit{TIC 159834975 (= HD 203006, A2pCrEuSr, V = 4.82)}\\

HD 203006 is a visual double with a close faint companion (2 mag fainter in the Hipparcos bandpass, separation of 0.1'',  \citealt{1997A&A...323L..53L}). 
A number of rotational periods have been published for HD 203006. \citet{1971PASP...83..474M} first reported probable photometric periods of 0.941\,d and 1.062\,d, then the latter was refined to 1.0609\,d by \citet{1973MitAG..32..252M}. 
The rotation period was found to be in fact about twice as long as 
previously thought 
(2.1219\,d, based on photometric and spectroscopic data; \citealt{1974A&A....32...21M}), which was then refined to $2.1215 \pm 0.0001$\,d \citep{1983A&A...118..289D}, although an automated period search using Hipparcos data yielded a period half as long \linebreak(1.0610\,d; \citealt{2011MNRAS.414.2602D}). We find a similar longer period in the \textit{TESS} data (2.122$\pm$0.004 d); it should be noted that it is not the strongest peak in the periodogram, as the first harmonic dominates the spectrum, hence the earlier confusion. 

HD 203006 has also been observed to vary in the ultraviolet, although there were not enough observations to establish a period \citep{1978A&A....66..187V}. Low-level variability was also observed in the near infrared with period of 2.1224\,d \citep{1998A&AS..129..463C}.
It was first observed to be magnetic by \citet{1958ApJS....3..141B}. Later undetected by \citet{1980ApJS...42..421B}, its field was eventually confirmed by \citet{1990MNRAS.247..606B}.\\





\noindent\textit{TIC 235007556 (= HD 221006, A0pSi, V = 5.68)}\\

HD 221006 was shown to vary photometrically with a period of $2.32 \pm 0.03$\,d \citep{1978IBVS.1391....1R}, which was later refined to $2.3148 \pm 0.0004$\,d \citep{1985A&AS...59..429M}. Spectroscopic variations also phase coherently with this period, as well as photometric variability in many filters, including in the infrared \citep{1995A&A...294..223L}. This star was also observed with Hipparcos, and an automated period search yielded a period of 2.3147\,d. The \textit{TESS} photometry of HD 221006 shows a similar period of 2.3119$\pm$0.0001\,d.
This star was detected to be magnetic by \citet{1993A&A...269..355B}.\\





\noindent\textit{TIC 237336864 (= HD 218495, A2pEuSr, V = 9.38)}\\

HD 218495 was found to host high frequency pulsations (P = 7.44 min; \citealt{1990IBVS.3509....1M, 1991MNRAS.250..666M}). No other variability period is reported in the literature for this roAp star prior to the \textit{TESS} observations. 

Although the periodogram is clearly dominated by the first harmonic (see Fig.~\ref{fig:astars}, the amplitude of the signal is clearly modulated on a period that is twice as long as the one that we would derive from the dominant peak), we determine a rotational period of 4.183$\pm$0.006\,d. 
This is similar to the value given by Cunha et al. (submitted), which offer the first determination of this star's rotational period (4.2006$\pm$0.0001\,d).

A first attempt to observe HD 218495's magnetic field did not yield a significant detection \citep{1997A&AS..124..475M}, but it has since been found to be magnetic \citep{2004A&A...415..685H}, a detection that was confirmed by \citet{2015A&A...583A.115B}.\\

\noindent\textit{TIC 262956098 (= HD 3988, A0pCrEuSr, V = 8.35)}\\

A double-lined spectroscopic binary with magnetically split lines ($B_d \sim 2.7$ kG; \citealt{2012MNRAS.420.2727E}
), HD 3988 does not show evidence of a companion in speckle interferometry \citep{1991MNRAS.248..411W}. It also does not vary with any known rotational or pulsational period \citep{1994MNRAS.271..129M}; similarly, no evidence of rotational modulation is found in the \textit{TESS} light curve.\\





\noindent\textit{TIC 277688819 (= HD 208217, A0pSrEuCr, V = 7.19)}\\

HD 208217 is an astrometric binary candidate based on proper motion measurements \citep{2007A&A...464..377F}.
HD 208217 was found to vary photometrically with a period of $8.35 \pm 0.10$\,d \citep{1983IBVS.2311....1M}. 
Its period was refined to $8.44475 \pm 0.00011$\,d \citep{1997A&A...320..497M} using both photometry and magnetic field modulus measurements derived from resolved line splitting 
(for the latter see also \citealt{1997A&AS..123..353M}; they mention this is a spectroscopic binary with a potentially long period, on the order of 2 years). We recover a slightly shorter period based on its \textit{TESS} light curve (8.317$\pm$0.001 d). 

HD 208217's magnetic field was detected and measured to provide a phase-resolved longitudinal field curve by \citet{2000A&A...359..213L}, with additional 
measurements published by \citet{2017A&A...601A..14M}.\\





\noindent\textit{TIC 278804454 (= HD 212385, A3pSrEuCr, V = 6.84)}\\

HD 212385 exhibits non-linear proper motion which makes it an astrometric binary candidate \citep{2005AJ....129.2420M, 2007A&A...464..377F}.
A period of $2.48 \pm 0.04$\,d was found in photometry \citep{1978IBVS.1391....1R}, and then refined to $2.5265 \pm 0.0015$\,d \citep{1985A&AS...59..429M}. We find a slightly shorter period in the \textit{TESS} data (2.5062$\pm$0.0002\,d). 
HD 212385 was discovered to be magnetic by \citet{2006AN....327..289H}, a detection confirmed by \citet{2006A&A...450..763K} and \citet{2015A&A...583A.115B}.\\
\\





\noindent\textit{TIC 279573219 (= HD 54118, A0pSi, V = 5.17)}\\

Identified as an astrometric binary candidate due to non-linear proper motion \citep{2005AJ....129.2420M, 2007A&A...464..377F}, HD 54118 was later found to be a spectroscopic binary by \citet{2012A&A...542A.116A}, though its companion is not well characterized. 
Its optical light curve was found to vary on a period of $3.275 \pm 0.015$\,d \citep{1981IBVS.2004....1M}, a period which was refined a few times using ground-based photometry \citep{1985A&AS...59..429M, 1993A&AS...97..501C}, most recently to $3.27535 \pm 0.00010$\,d \citep{1994A&A...281...73M}. This star was also observed with Hipparcos, and an automated period search yielded a rotational period of 3.2749\,d \citep{2011MNRAS.414.2602D}.

Found to be magnetic with $P_\textrm{rot} = 3.2 \pm 0.1$\,d \citep{1975PASP...87..961B}, its field detection was confirmed by \citet{1993A&A...269..355B} using phase resolved observations; they also refined the period to $3.27533 \pm 0.00020$\,d, which is consistent with the photometric period. There was also a further follow-up spectroscopic observation by \citet{1997MNRAS.291..658D}.
The \textit{TESS} light curve reveals a period of 3.2759$\pm$0.0002\,d, consistent with the previous results.\\

\noindent\textit{TIC 280051011 (= HD 18610, A2pCrEuSr, V = 8.14)}\\

Zeeman splitting in the spectra of HD 18610 led to discovery of a magnetic field with a modulus of about 5.7 kG \citep{2003A&A...402..729S}. This star has no rotational or pulsational periods reported in the literature \citep{1994MNRAS.271..129M}, and no signature of rotation is detected in the \textit{TESS} data.\\

\noindent\textit{TIC 281668790 (= HD 3980, A7pSrEuCr, V = 5.70)}\\

HD 3980 is a known visual double (\citealt{1955AnAp...18..379K, 2014MNRAS.437.1216D}; its companion is 2.89 mag fainter in the $K$ band and separated by $\sim$13'').
The first photometric period determined for HD 3980 was 0.4\,d \citep{1976MitAG..38..177M}. Since then, different photometric periods have been reported: 2.13\,d or 0.68\,d \citep{1979A&A....77..366R}, and later $3.9516 \pm 0.0003$\,d  \citep{1980A&A....81..323M}. The latter study also took into account magnetic measurements based on Zeeman line broadening/splitting. 
Photometric variability in the infrared ($J$, $H$ and $K$ bands) is also found to phase coherently with the longer optical period \citep{1991A&A...248..179C, 1998A&AS..129..463C}. Hipparcos data reveal, as a result of an automated search, a potential 1.1628\,d period \citep{2011MNRAS.414.2602D}, but this value does not agree with any other study.

We find a period in the \textit{TESS} light curve that is similar to the values obtained from earlier ground-based observations (3.951$\pm$0.003\,d); it should be noted that it is not the strongest peak in the periodogram, as the first harmonic dominates the spectrum. 
The detection of a magnetic field at the surface of HD 3980 has been reported by \citet{2006AN....327..289H} and \citet{2015A&A...583A.115B}, supporting the earlier Zeeman line broadening measurements.\\






\noindent\textit{TIC 348717688 (= HD 19918, A5pSrEuCr, V = 9.35)}\\

HD 19918 is a well-known roAp star discovered in the Cape survey (\citealt{1991IBVS.3553....1M, 1994MNRAS.271..118M,1995MNRAS.276.1435M}; see also Cunha et al. submitted) with no visual companion \citep{2012A&A...545A..38S}.
A marginal field detection \citep{1997A&AS..124..475M} was later confirmed by \citet{2006A&A...450..763K} (also see \citealt{2006AN....327..289H, 2015A&A...583A.115B}), and investigated through line broadening as well \citep{2007A&A...473..907R}.
The \textit{TESS} light curve does not yield any frequency peak compatible with rotational modulation.\\





\noindent\textit{TIC 358467700 (= HD 65712, A0pSiCr, V = 9.35)}\\

HD 65712 is a member of NGC 2516 \citep{1975PASP...87..721S}, despite being identified as a possible high-velocity star by \citet{1983A&A...127....1J}. Its membership in the cluster was confirmed by \citet{2007A&A...470..685L}. 

HD 65712 was found to have a 1.943$\pm$0.001-d photometric rotation period by \citet{2004BaltA..13..597W}, although a shorter period of 1.88\,d was found by \citet{2011A&A...525A..16P}. We find a period in the \textit{TESS} photometry of this star (1.9460$\pm$0.0002\,d) that appears to be more consistent with the earlier determination of \citet{2004BaltA..13..597W}; hence, we choose to report that value in Table~\ref{tab:targets}.
A magnetic field was detected by \citet{2006A&A...450..777B} (and confirmed by \citealt{2015A&A...583A.115B}).\\

\noindent\textit{TIC 410451752 (= HD 66318, A0pEuCrSr, V = 9.56)}\\

HD 66318 is a probable member of NGC 2516 \citep{1955ApJ...121..628C, 2008AJ....136..118F}. 
No previous study of this star has discovered photometric variability \citep{1982IBVS.2103....1N, 1987A&AS...69..371N, 1989A&AS...78...25D}.
A strong magnetic field was detected by \citet{2003A&A...403..645B} and later confirmed by \citet{2006A&A...450..777B} and \citet{2015A&A...583A.115B}, with a significant discrepancy observed between the field strengths measured from hydrogen and metal lines \citep{2014A&A...572A.113L}. These authors do not find spectral variability, suggesting a very long rotation period (potentially on the order of years). This is also consistent with their finding that the projected rotational velocity is very low as determined from high-resolution UVES spectra, contradicting previous findings that $v \sin i$ = 30 km s$^{-1}$ \citep{1972A&A....21..373D, 1981ApJ...244..221W}; this indicates that the viewing angle might simply not allow us to detect significant rotational modulation.

Similarly, we do not detect a convincing rotational period in the \textit{TESS} photometry; as such, we reiterate the conclusion of \citet{2017A&A...601A..14M} that more observations are required to characterize the long-term variability of this star. However, there appears to be a weak potential rotational peak at 1.3 d$^{-1}$ (and its first harmonic)
; given that the pixels from which this star's signal is measured are highly contaminated (contamination ratio of 0.815 according to the TIC), it is plausible that the flux of nearby stars is within the light curves of HD~66318. A similar conclusion is reached by Cunha et al. (submitted), who report a period of $0.77688\pm0.00052$\,d. 


\section{Discussion and conclusions}\label{sec:concl}

This paper introduces the MOBSTER collaboration, a group consisting of both observers and theorists with the aim of using \textit{TESS} data to further the study and characterization of magnetic OBA stars and to discover new magnetic stars out of a photometrically pre-selected sample of targets. Upcoming studies will include the characterization of specific magnetic objects of interest via spectropolarimetry and will confront analytic models with observations.

Twelve out of the nineteen known magnetic B and A stars observed in the first two sectors of \textit{TESS} show periodograms 
displaying a main frequency peak and at least one harmonic -- this is a characteristic signature of rotational modulation (e.g. \citealt{2018A&A...616A..77B}), understood to be due to the presence of an oblique dipolar magnetic field -- 
with an additional star exhibiting only a fundamental peak, but whose light curve still appears to be rotationally modulated
. 
Other targets showing this behaviour in the periodogram of their \textit{TESS} light-curves should be considered as promising magnetic candidates. Such candidates have been identified for OB stars in sectors 1 and 2 by \citet{2019ApJ...872L...9P} and for A stars in sectors 1 to 4 by Sikora et al. (submitted). They are prime targets for future spectropolarimetric observations. This identification is a crucial first step to increase the sample size of magnetic OBA stars (especially for the earliest spectral types) since spectropolarimetry is an expensive observational technique. Large spectropolarimetric surveys have uncovered a number of such stars, but the efficiency limit of these (essentially magnitude limited) surveys has been reached. Therefore, photometrically pre-selecting strong candidates among a fainter sample of stars with a high expected detection rate (e.g. \citealt{2018MNRAS.478.2777B}) constitutes our best possible strategy moving forward, and this is uniquely enabled by the high quality light curves obtained with \textit{TESS}.

We have refined the period determinations for thirteen targets, and in some cases, compared to published values based both on photometric studies and other types of observations. We also present phase folded light curves for these thirteen stars in the Appendix (Fig.~\ref{fig:app}); these show a wide variety of morphologies and nicely illustrate the exquisite quality of the \textit{TESS} observations. We find our results to be consistent with the literature values overall, and in particular, unsurprisingly, with the values derived by Cunha et al. (submitted) using the same dataset. In a few cases (especially in the case of HD 223640), there is an apparently significant departure between our values and the latter; this is likely due to the different methodologies employed to calculate the rotational periods. 
These results also illustrate the immense potential of \textit{TESS}, compared to other large space-based surveys, as we detect rotational modulation and derive the rotational period in three stars (HD 65987, HD 208217 and HD 212385) that were observed by Hipparcos, but for which no period was recovered.

As for the six 
stars that did not show significant low frequency peaks in their periodograms (HD 213637, HD 217522, HD 3988, HD 18610, HD 19918 and HD 66318), we are not able to conclude anything about their rotational periods. Half of them (HD 213637, HD 217522 and HD 19918) are known to exhibit rapid oscillations, identifying them as roAp stars, while the last one is likely too heavily contaminated for its rotational modulation to be significantly detected (although a weak signal with a period of about 0.8\,d might be present). The most useful constraint to evaluate these stars' rotational periods would be to obtain phase-resolved magnetic measurements, although it is possible that the lack of apparent rotational modulation is due to an unfavourable alignment of the rotational and magnetic axes with respect to the line of sight.

\section*{Acknowledgements}

This paper includes data collected by the TESS mission. Funding for the TESS mission is provided by the NASA Explorer Program. Funding for the TESS Asteroseismic Science Operations Centre is provided by the Danish National Research Foundation (Grant agreement no.: DNRF106), ESA PRODEX (PEA 4000119301) and Stellar Astrophysics Centre (SAC) at Aarhus University. We thank the TESS and TASC/TASOC teams for their support of the present work. This research has made use of the SIMBAD database, operated at CDS, Strasbourg, France. Some of the data presented in this paper were obtained from the Mikulski Archive for Space Telescopes (MAST). STScI is operated by the Association of Universities for Research in Astronomy, Inc., under NASA contract NAS5-2655.

The authors are grateful to C.L. Fletcher for her linguistic assistance. We also thank the referee (G. Mathys) who provided useful comments and contributed to improving this paper.

ADU, VK, CCL and GAW acknowledge the support of the Natural Science and Engineering Research Council of Canada (NSERC). The research leading to these results received funding from the European Research Council (ERC) under the European Union's Horizon 2020 research and innovation program (grant agreement No. 670519: MAMSIE). VP acknowledges support from the National Science Foundation under Grant No. 1747658. SC and GH gratefully acknowledge funding through the Polish NCN grant 2015/18/A/ST9/00578. JL-B acknowledges support from FAPESP (grant 2017/23731-1). MES acknowledges the financial support provided by the Annie Jump Cannon Fellowship, supported by the University of Delaware and endowed by the Mount Cuba Astronomical Observatory. AuD acknowledges support from NASA through Chandra Award number 
TM7-18001X issued by the Chandra X-ray Observatory Center, which is 
operated by the Smithsonian Astrophysical Observatory for and on behalf 
of NASA under contract NAS8-03060.




\bibliographystyle{mnras}

\bibliography{database}



\appendix

\section{Phased light curves}\label{sec:app}

In this section, we present phase folded light curves for each star showing rotational modulation (with overlaid binned light curves using 20 phase bins over the full rotational cycle).

\begin{figure*}
	\begin{subfigure}{0.32\linewidth}
	\includegraphics[width=\linewidth]{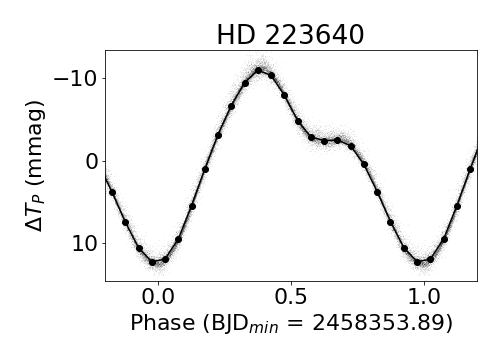}
	\end{subfigure}
	\begin{subfigure}{0.32\linewidth}
	\includegraphics[width=\linewidth]{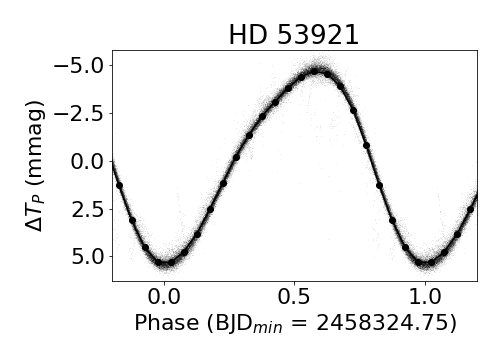}
	\end{subfigure}
	\begin{subfigure}{0.32\linewidth}
	\includegraphics[width=\linewidth]{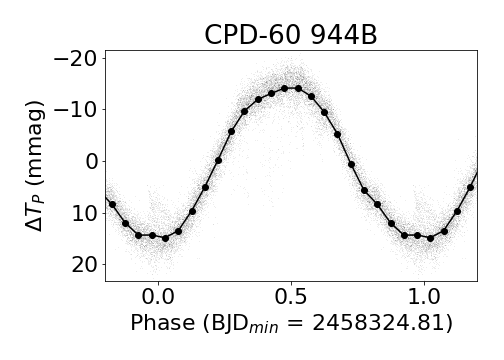}
	\end{subfigure}
	\begin{subfigure}{0.32\linewidth}
	\includegraphics[width=\linewidth]{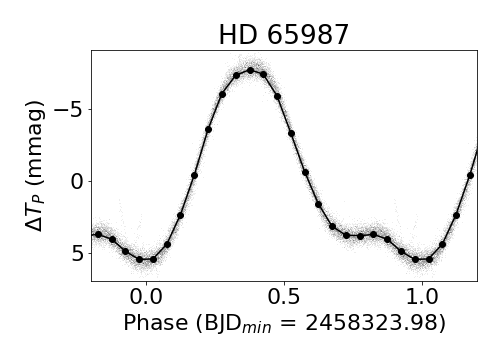}
	\end{subfigure}
	\begin{subfigure}{0.32\linewidth}
	\includegraphics[width=\linewidth]{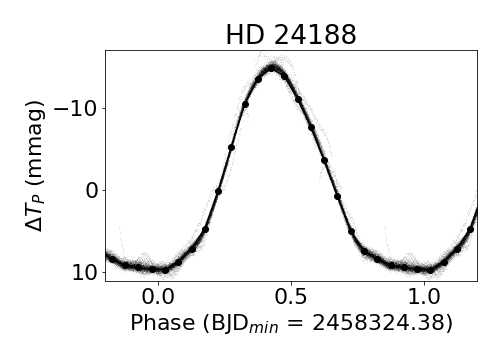}
	\end{subfigure}
	\begin{subfigure}{0.32\linewidth}
	\includegraphics[width=\linewidth]{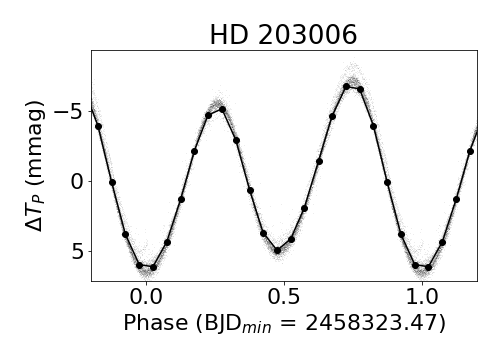}
	\end{subfigure}
	\begin{subfigure}{0.32\linewidth}
	\includegraphics[width=\linewidth]{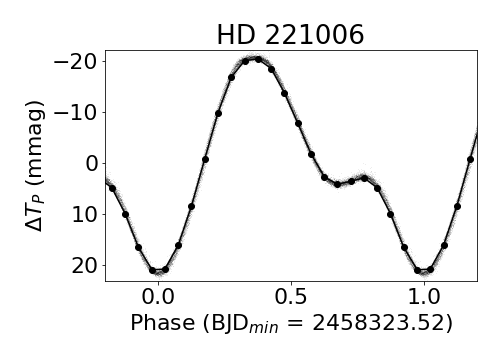}
	\end{subfigure}
	\begin{subfigure}{0.32\linewidth}
	\includegraphics[width=\linewidth]{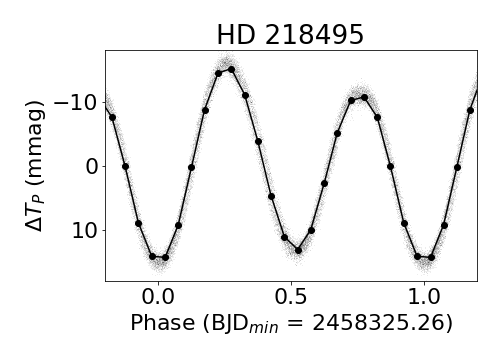}
	\end{subfigure}
	\begin{subfigure}{0.32\linewidth}
	\includegraphics[width=\linewidth]{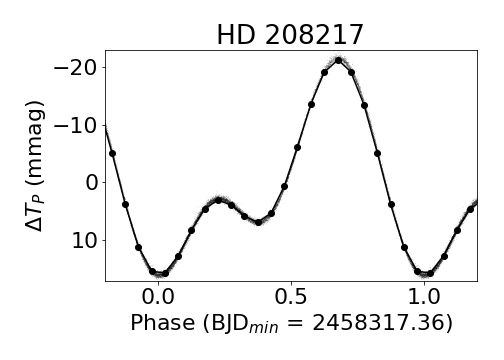}
	\end{subfigure}
	\begin{subfigure}{0.32\linewidth}
	\includegraphics[width=\linewidth]{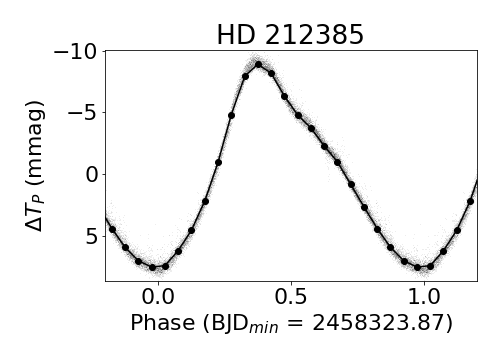}
	\end{subfigure}
	\begin{subfigure}{0.32\linewidth}
	\includegraphics[width=\linewidth]{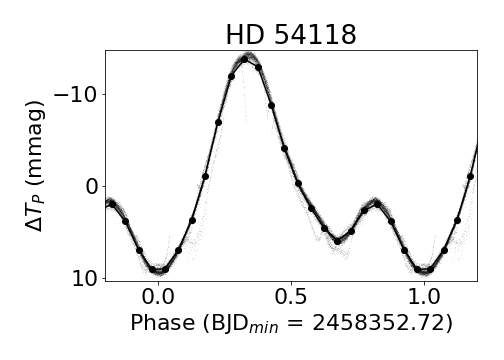}
	\end{subfigure}
	\begin{subfigure}{0.32\linewidth}
	\includegraphics[width=\linewidth]{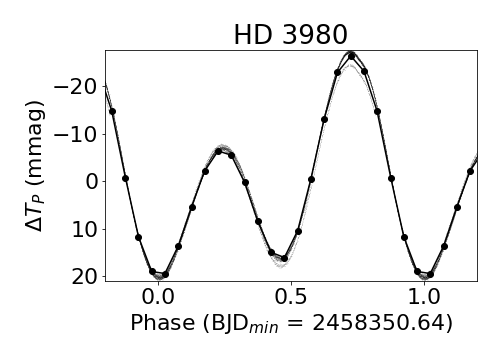}
	\end{subfigure}
	\begin{subfigure}{0.32\linewidth}
	\includegraphics[width=\linewidth]{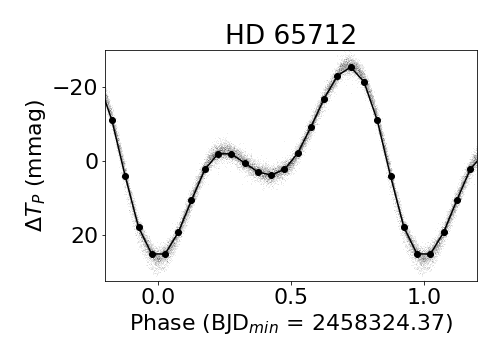}
	\end{subfigure}
    \caption{Phase folded light curves (with phase 0 corresponding approximately to minimum light; the time of minimum light used to phase the light curve is indicated on the x-axis of each plot for reference) for the thirteen stars showing potential rotational modulation. The light grey points correspond to individual measurements, and the larger black points connected by a line correspond to binned data (with bins of 0.05 in phase). For some stars, we see streaks of outliers; these are due to artefacts which could not be completely detrended. We find that the light curves have diverse morphologies, with at least nine out of the thirteen showing signs of double-wave variations, and all of them having an asymmetric profile.}
    \label{fig:app}
\end{figure*}


\bsp	
\label{lastpage}
\end{document}